\documentclass[final]{aipproc}

\layoutstyle{6x9}

\begin{document}

\title{Charge symmetry breaking and 
the \boldmath$dd\to\alpha\pi^0$ reaction --- 
recent theoretical developments}

\author{{Anders G{\aa}rdestig}}{
  address={Ohio University, Athens, Ohio, U.S.A.}
  ,email=anders@phy.ohiou.edu
}

\begin{abstract}
The status of the theoretical effort to calculate charge-symmetry-breaking pion 
production reactions will be presented.
The main emphasis will be on the $dd\to\alpha\pi^0$ reaction, especially in relation
to the recent IUCF experiment. 
Suggestions for future theoretical and experimental work will be made.
\end{abstract}

\maketitle

\section{Introduction}
This presentation is based mainly on~\cite{CSB1}, work done in collaboration with 
C.~J.~Horowitz, A.~Nogga, A.~C.~Fonseca, 
C.~Hanhart, G.~A.~Miller, J.~A.~Niskanen, and U.~van Kolck. 

The Lagrangian of quantum chromodynamics (QCD) is almost symmetric under the
exchange of the up and down quarks, which is called charge symmetry (CS).
However, since the quark masses are different, this symmetry is broken, charge 
symmetry breaking (CSB).
For hadrons, CS implies the invariance of the strong interaction under,
e.g., the exchange of the proton and neutron. 

Experimental evidence for CSB has been demonstrated (before 2002) in $\rho^0$-$\omega$ 
mixing, the neutron-proton and other hadron mass splittings, the binding-energy 
difference of mirror nuclei such as $^3{\rm H}$ and $^3{\rm He}$, the different 
scattering lengths of elastic $nn$ and $pp$ scattering, and in the difference between 
the proton and neutron analyzing powers of elastic $np$ scattering. 
A review about CSB and related issues can be found in Ref.~\cite{MNS}.

The $n$-$p$ mass difference can be split into two parts, 
$m_n-m_p=\delta M+\bar\delta M=1.29$~MeV/$c^2$, where $\delta M$ is due to the 
strong interaction and stems from the up-down quark mass difference, while
$\bar\delta M$ is due to electromagnetic effects. 
The latter is known to be negative and various models gives typically 
$\bar\delta M\sim -1$~MeV/$c^2$~\cite{vKNM}, but its precise value is not known.
How can we determine these constants?
The Lagrangian of chiral perturbation theory ($\chi$PT) contains two terms 
parametrized by $\delta M$ and $\bar\delta M$:
\begin{equation}
  \mathcal{L}_{CSB} = \frac{\delta M}{2}N^\dagger\left(\tau_3-
	  \frac{\pi_3\tau\cdot\pi}{2f_\pi^2}\right)N+
        \frac{\bar\delta M}{2}N^\dagger\left(\tau_3+
       \frac{\pi_3\tau\cdot\pi-\pi^2\tau_3}{2f_\pi^2}\right)N.
\end{equation}
The pion terms may give a different combination of $\delta M$ and $\bar\delta M$ than
for $m_n-m_p$, which makes it possible to use reactions involving pions to extract 
$\delta M$ and $\bar\delta M$. 
A first suitable choice is the CSB forward-backward asymmetry of $np\to d\pi^0$, where 
originally a negative value $A_{\rm fb}=-0.28\%$ was predicted from
$\eta$-$\pi^0$ mixing~\cite{nis99}.
When the $\delta M$ and $\bar\delta M$ terms were included the sign changed and 
$A_{\rm fb}=0.23\%$--$0.60\%$~\cite{vKNM}.
This latter prediction is of the same sign and magnitude as the subsequent 
experimental result from TRIUMF:
$A_{\rm fb}=[17.2\pm8({\rm stat})\pm5.5({\rm sys})]\times 10^{-4}$~\cite{Allena}.

Another possiblity is the $dd\to\alpha\pi^0$ reaction, which is CSB since the 
deuteron and $\alpha$ particle are self-conjugate under CS, while the pion wave 
function changes sign. 
Thus this reaction could not occur if CS were conserved.
The cross section is proportional to the square of the CS amplitude, which is unique 
since all other CSB observables involve interferences with CS amplitudes.
A long series of experimental work, culminating with the efforts at the Saturne 
accelerator at Saclay, lead only to decreasingly small upper limits~\cite{Banaigs2}.
Finally, the Saturne group claimed to see $dd\to\alpha\pi^0$ at 
$T_d=1.1$~GeV~\cite{Goldzahl}. 
This was refuted by members of the same collaboration who argued that the
\mbox{$dd\to\alpha\gamma\gamma$} reaction was responsible for the signal~\cite{FP}.
The importance of this background was later confirmed by calculation of the 
double-radiative capture~\cite{DFGW}, using a model based on a very successful 
treatment of $dd\to\alpha\pi\pi$ at similar energies~\cite{ABC}.
Thus, the claimed signal is most likely a misinterpretation of a heavily-cut smooth 
$dd\to\alpha\gamma\gamma$ background~\cite{DFGW}.
This realization influenced the design of the IUCF experiment, which last year
reported a very convincing $dd\to\alpha\pi^0$ signal near threshold 
($\sigma=12.7\pm2.2$~pb at $T_d=228.5$~MeV and $15.1\pm3.1$~pb at 231.8~MeV), 
superimposed on a smooth $dd\to\alpha\gamma\gamma$ background~\cite{IUCFCSB}. 
This background is roughly a factor two larger than calculations based on 
Ref.~\cite{DFGW}, but has the expected shape.
The data are consistent with the pion being produced in an $s$-wave,
as expected from the proximity of the threshold ($T_d=225.6$~MeV).

It is hoped that these two experimental results, when analyzed in a consistent
theoretical framework, can be used to 
extract the values of $\delta M$ and $\bar\delta M$~\cite{vKNM}.
Our aim here is to provide the first study of CSB in the near threshold 
$dd\to\alpha\pi^0$ reaction using chiral techniques.
Previous models of $dd\to\alpha\pi^0$ have been at higher energies and have
used $\eta$-$\pi^0$ mixing only~\cite{Cheung82,cp86}.
We are guided in our quest by general symmetry considerations, relying on chiral
power counting to give a list of possible amplitudes, and trusting that realistic 
wave functions can be constructed for the four-body states.

\section{Power counting}
\begin{figure}
  \includegraphics[width=60mm]{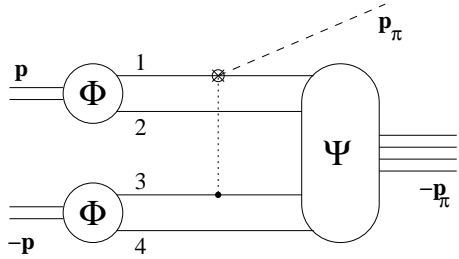}
  \caption{Feynman diagram for leading order pion production in the $dd\to\alpha\pi^0$ 
    reaction (simplified model), indicating the labeling of nucleons and defining 
    basic kinematic variables.
    The cross indicates the occurrence of CSB and the dot a leading-order CS vertex.}
\label{fig:ddapi0}
\end{figure}
The relevant scales are the typical momentum for pion production
($\chi=p/M\approx\sqrt{m_\pi/M}$) ordering the strong diagrams and the fine structure 
constant ($\alpha\approx1/137$) ordering electromagnetic effects.
Starting with the strong diagrams, at leading order (LO) there is only the
$\pi$ rescattering diagram of Fig.~\ref{fig:ddapi0}, which is driven by 
$\delta M$ and $\bar\delta M$.

\begin{figure*}
\includegraphics[width=88mm]{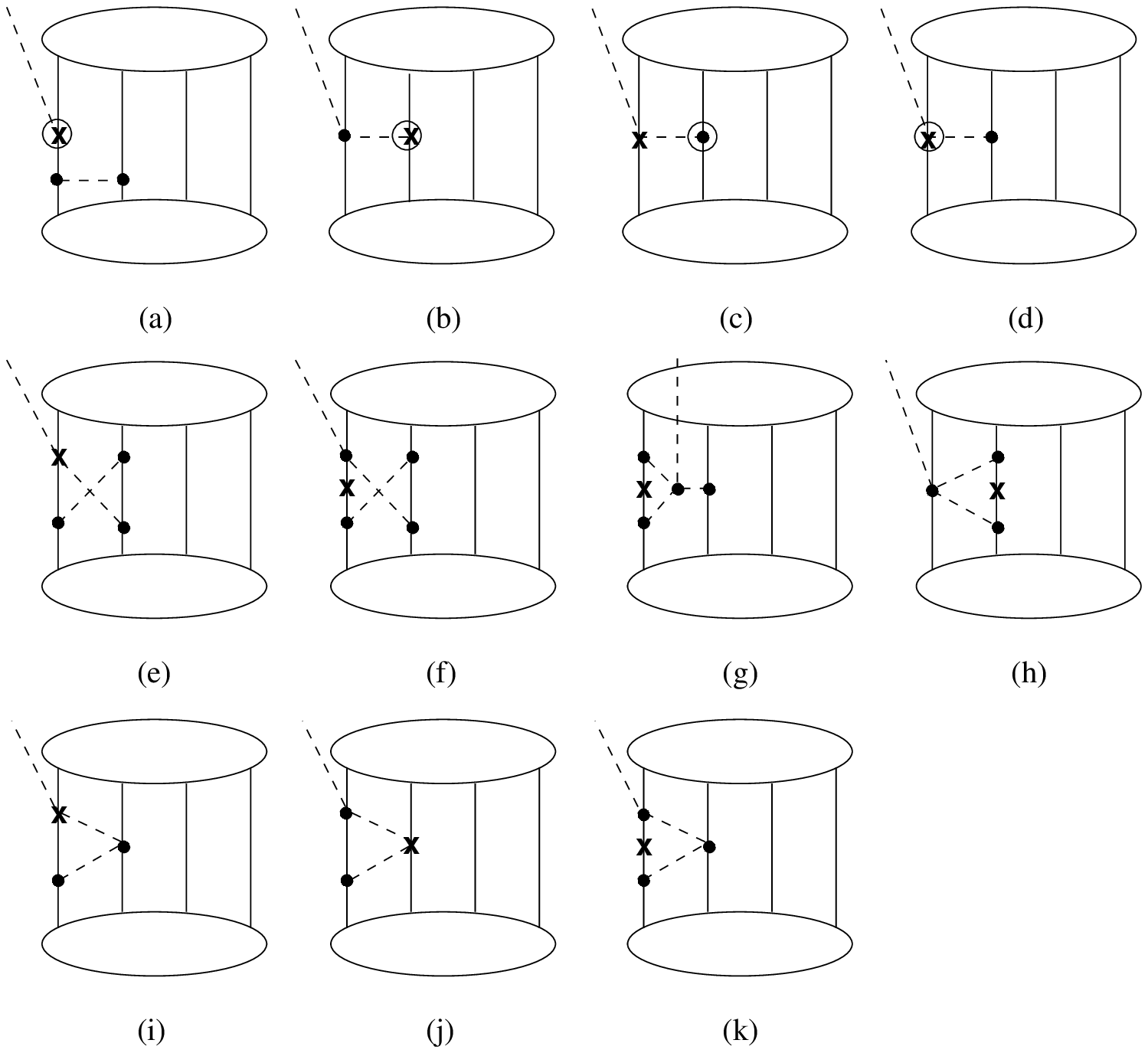}\hspace{10mm}%
\parbox[b]{38mm}{\includegraphics[width=41.8mm]{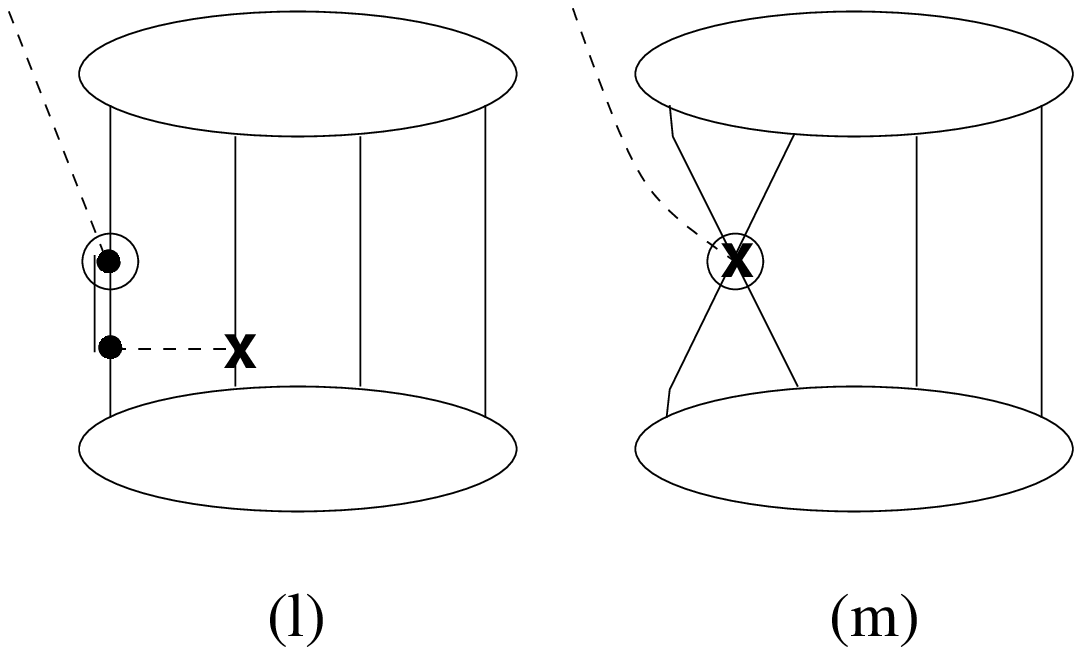}\\[12.1ex]}
\caption{Left: NNLO diagrams with strong CSB [(a)--(k)].
  Encircled vertices are sub-leading. Not all possible time orderings are shown.
Right: Some typical higher-order diagrams with strong CSB [(l) and (m)]. 
    A double line represents a $\Delta$-isobar.
    Diagram (l) appears at N$^3$LO whereas diagram (m) is a N$^4$LO contribution. }
\label{fig:NNLOeps}
\end{figure*}

At next-to-next-to-leading order (NNLO) there are several diagrams as displayed in
Fig.~\ref{fig:NNLOeps}.
The first (a) (one-body) term is parametrized by $\eta$-$\pi^0$ mixing (resonance 
saturation) $\beta_1=g_{\eta}\langle\pi^0|H|\eta\rangle/m_{\eta}^2$, unfortunately 
with large uncertainties in $g_\eta$ and $\langle\pi^0|H|\eta\rangle$.
The pion exchanges (b)--(d) are smaller.
We have not yet included the pion loops (e)--(k).

The N$^x$LO diagram of Fig.~\ref{fig:NNLOeps} are in $\chi$PT given by low-energy
constants (LECs) that should eventually be fitted to data. 
Since the LECs are not known we chose to model diagram (b) by assuming resonance
saturation, i.e., heavy-meson ($\sigma$, $\rho$, $\omega$) exchanges.
Also $\rho$-$\omega$ mixing and hard photon exchanges contribute.
This is very similar to the situation in $pp\to pp\pi^0$ where short-range
mechanisms were also needed~\cite{leeriska,chucketal}.
The CSB parameter is again $\beta_1$. 
The $\Delta$-excitations (l) has not been calculated.

The electromagnetic leading order causes CSB in the wave functions.
Higher orders contain CSB pion production.
These amplitudes have not yet been included in our model.

\section{Simplified model}
As a first step toward a full model we chose to use simplified (Gaussian) wave 
functions, without initial state distortions, to gain insight into the 
problem~\cite{CSB1}.
The diagram for the simplified model is given in Fig.~\ref{fig:ddapi0}.
The deuterons need to be slowed down (the c.m.\ momentum is $p\sim460$~MeV/$c$)
in order to produce a pion and $\alpha$ particle at rest. 
Thus we would like to have momentum sharing between the deuterons so that it does not 
push the nucleons far out in the high-momentum tail of the $\alpha$-particle wave 
function.

With spatially symmetric wave functions the spin-isospin parts are anti-symmetrized:
\begin{eqnarray*}
     |d_{12}\rangle & = & (12)_1[12]_0,\hspace{5mm}|d_{34}\rangle = (34)_1[34]_0 \\
     |\alpha\rangle & = & \frac{1}{\sqrt{2}}\left\{
     ((12)_1(34)_1)_0[[12]_0[34]_0]_0-((12)_0(34)_0)_0[[12]_1[34]_1]_0 \right\}
\end{eqnarray*}
where $(i,j)_s$ ($[i,j]_T$) are the spin (isospin) Clebsch-Gordan 
couplings, with magnetic quantum numbers suppressed, for nucleons, or nucleon 
pairs, $i$ and $j$ coupling to spin $s$ (isospin $T$). 
Amplitudes easily match $|dd\rangle$ with the first part of $|\alpha\rangle$, 
while the second term is matched only if both spin and isospin of
both nucleon pairs are flipped simultaneously.
Thus symmetries decide about matching, but do they support momentum sharing?

The symmetries forbid LO pion exchange between the deuterons and only the NNLO recoil 
piece $(-m_\pi/M)p_N$ survives [Fig.~\ref{fig:NNLOeps}(c)].
Thus the LO is strongly suppressed.

The NNLO one-body cannot allow for momentum sharing and only its recoil 
part can contribute to $s$-wave production, but it is supported by the symmetries and 
adds coherently over all nucleons. It is quite small, of the order 1~pb.

The short range N$^4$LO amplitudes survive the symmetries, allow for momentum sharing, 
and add coherently over all nucleons and with each other, dominating the cross section 
in our simplified model, as demonstrated in Table~\ref{tab:XS}.

\section{Results}
The diagrams calculated so far are summarized in Table~\ref{tab:XS}, where a 
comparison is also made with the plane wave impulse approximation (PWIA)
using realistic wave functions, (CD-Bonn+Urbana or AV18+Urbana).
The pion rescattering recovers somewhat, since new symmetries possible 
with realistic wave functions allow for momentum sharing. 
The one-body term increases dramatically since it does not have any momentum 
sharing and the high-$p$ tail of the $\alpha$ wave function is no longer a tiny
Gaussian.

\begin{table}
\begin{tabular}{c|ccc}
\tablehead{1}{c}{b}{Operator\\(mech.)} & \tablehead{1}{c}{b}{SWF\\
{[pb]}} & \tablehead{1}{c}{b}{PWIA (CD-Bonn)\\
{[pb]}} & \tablehead{1}{c}{b}{PWIA (AV18)\\
{[pb]}} \\ \hline
$\pi$($\delta,\bar\delta$) & 0.011 & 1.619  & 1.384   \\
1($\eta$-$\pi$)            & 0.688 & 12.561 & 10.341  \\
$\sigma$($\eta$-$\pi$)     & 0.187 & 0.826  & 0.467   \\
$\omega$($\eta$-$\pi$)     & 0.404 & 1.084  & 0.759   \\
$\rho$($\eta$-$\pi$)       & 0.081 & 0.185  & 0.092   \\
$\rho$-$\omega$            & 1.645 & 4.191  & 2.647   \\
$\gamma$(EM)               & 1.486 &        &         \\
CSC 1(EM in wf)            &       & 0.095  & 0.080   \\
total                      & 22.955 & 79.832 & 46.508 \\
no $\eta$-$\pi$            & 1.925  & 9.065  & 3.049  \\ \hline 
Expt. & \multicolumn{3}{c}{$12.7\pm2.2$}
\end{tabular}
\caption{Cross sections evaluated for $dd\to\alpha\pi^0$ 
at the lowest IUCF energy, for simplified wave functions (SWF) and realistic
(PWIA), either CD-Bonn or AV18. 
The experimental cross section is also given.}
\label{tab:XS}
\end{table}

While the simplified model gets quite close to the experimental value, the realistic 
wave functions, without distortion, overshoot considerably.
There are several reasons why this may happen.
Firstly, this can be because the initial state interactions are not included yet.
Secondly, the ignored diagrams (e.g., the pion loops at NNLO) may cause destructive
interferences.
Thirdly, since the strength is mainly driven by the one-body term, the resonance 
saturation procedure may not be accurate?
To test this we turned off the $\eta$-$\pi$ mixing, which gave more reasonable cross 
sections.
Another concern is the difference between the CD-Bonn and AV18 calculations.
However, our dependence on potential choices can be assessed only when the
full model is in place.

\section{Conclusion and outlook}
We find that, at least in our simplified model, the LO diagrams are suppressed,
while subleading amplitudes are important and dominate the cross section.
Thus, these short-range terms may be important for $np\to d\pi^0$
and should be evaluated there as well.

We are working on the completion of our model, especially regarding the $dd$ initial
state. 
Polarized $dd\to dd$ data currently being analyzed at IUCF will be used to test 
the scattering wave functions.
The remaining diagrams, the NNLO pion loops, photon two- and three-body terms, 
and possible $\Delta$ ($p$-wave pion) contributions, will be included.
The dependence on the choice of wave function and potentials can then be tested.

With realistic bound state wave functions the cross section is overestimated, 
which may reflect our ignorance of the unknown LECs.
To address this issue, new experiment are necessary, e.g., $np\to d \pi^0$ and 
unpolarized or polarized $dd\to\alpha\pi^0$, with angular distributions and 
possibly at higher energies.
As a $dd\to\alpha\pi^0$ experiment is being planned at COSY, hopefully 
new exciting data will be available within a few years.

This work was supported in part by the DOE grants DE-FG02-93ER40756 and 
DE-FG02-02ER41218.

\bibliographystyle{aipproc}

\end{document}